\documentclass[conference,compsoc]{IEEEtran}
\IEEEoverridecommandlockouts

\pdfoutput=1

\usepackage{graphicx}
\usepackage[table,xcdraw]{xcolor} 
\usepackage{multirow}
\usepackage{tabularx,booktabs}
\usepackage{graphicx}
\usepackage{balance}  
\usepackage{latexsym,fancyhdr,url}

\usepackage{
  color,
  xcolor,
  float,
  epsfig,
  wrapfig,
  graphics,
  graphicx,
  subcaption
}

\usepackage[sort, square,numbers]{natbib}
\usepackage{hyperref}
\hypersetup{hidelinks}

\usepackage{url}
\def\BibTeX{{\rm B\kern-.05em{\sc i\kern-.025em b}\kern-.08em
    T\kern-.1667em\lower.7ex\hbox{E}\kern-.125emX}}

\usepackage{amsmath,amssymb,amsfonts}
\usepackage{textcomp}
\usepackage{xspace}
\usepackage{enumitem}

\usepackage{algorithmic}

\usepackage{wrapfig}
\usepackage{titlesec}

\usepackage{fancyhdr}
\usepackage{hyperref}
\hypersetup{
    colorlinks=false,
    linkcolor=black,
    filecolor=black,      
    urlcolor=black,
    pdftitle={Overleaf Example},
    pdfpagemode=FullScreen,
    }
\newcommand{\bhref}[3][teal]{\href{#2}{\color{#1}{#3}}}%

\begin{document}
\title{SoK: Safer Digital-Safety Research Involving At-Risk Users\\}


\author{\vspace*{4em}}


\author{
\centering
\IEEEauthorblockN{
Rosanna Bellini\IEEEauthorrefmark{1} 
\hspace*{1em}
Emily Tseng\IEEEauthorrefmark{1}
\hspace*{1em}
Noel Warford\IEEEauthorrefmark{2}
\hspace*{1em}
Alaa Daffalla\IEEEauthorrefmark{1}}
\IEEEauthorblockN{
Tara Matthews\IEEEauthorrefmark{3}
\hspace*{1em}
Sunny Consolvo\IEEEauthorrefmark{3}
\hspace*{1em}
Jill Palzkill Woelfer\IEEEauthorrefmark{4}
\hspace*{1em}
Patrick Gage Kelley\IEEEauthorrefmark{3}}
\IEEEauthorblockN{
Michelle L. Mazurek\IEEEauthorrefmark{2}
\hspace*{1em}
Dana Cuomo\IEEEauthorrefmark{5}\hspace*{1em}
Nicola Dell\IEEEauthorrefmark{1}
\hspace*{1em}
Thomas Ristenpart\IEEEauthorrefmark{1}
}
\vspace*{6pt}
\IEEEauthorblockN{\IEEEauthorrefmark{1}Cornell Tech 
\hspace*{2em}
\IEEEauthorrefmark{2}University of Maryland 
\hspace*{2em}
\IEEEauthorrefmark{3}Google
\hspace*{2em}
\IEEEauthorrefmark{4}JumpCloud
\hspace*{2em}
\IEEEauthorrefmark{5}Lafayette College}
}

\maketitle

\renewcommand{\paragraph}[1]{\vspace*{6pt}\noindent\textbf{#1}\;}

\newcommand{\CC}[1]{\cellcolor{gray!#1}}

\newcommand{\secref}[1]{Section~\ref{#1}}
\newcommand{\figref}[1]{Figure~\ref{#1}}
\newcommand{\tabref}[1]{Table~\ref{#1}}
\newcommand{\apref}[1]{Appendix~\ref{#1}}

\newcommand{\fixme}[1]{\ifnum\authnote=1{\textcolor{red}{[FIXME: #1]}}\fi}
\newcommand{\better}[1]{\ifnum\authnote=1{\textcolor{violet}{[BetterWord: #1]}}\fi}
\newcommand{\todo}[1]{\ifnum\authnote=1{\textcolor{red}{[TODO: #1]}}\fi}

\newcommand{\new}[1]{\textcolor{blue}{#1}}

\newcounter{mynote}[section]
\newcommand{\notecolor}{blue}
\newcommand{\thenote}{\thesection.\arabic{mynote}}

\newcommand{\myemph}[1]{\vspace*{3pt}\emph{\textbf{#1}}\hspace*{3pt}}
\newcommand{\ronote}[1]{\textcolor{red}{$\ll${RB~\thenote:} {\sf #1}$\gg$}}
\newcommand{\emnote}[1]{ \textcolor{orange}{$\ll$ET~\thenote: {\sf #1}$\gg$}}

\newcommand\edit[1]{\textcolor{black}{#1}}
\newcommand\rep[1]{\textcolor{blue}{#1}}
\newcommand\change[1]{\textcolor{teal}{#1}}

\newcommand{\ds}{{\textcolor{black}{digital-safety}}\xspace}
\newcommand{\Ds}{{\textcolor{black}{Digital-safety}}\xspace}
\newcommand{\DS}{{\textcolor{black}{Digital Safety}}\xspace}
\newcommand{\dsnoun}{{\textcolor{black}{digital safety}}\xspace}

\newcommand{\paperNo}{\edit{{196}}\xspace}
\newcommand{\subsetNo}{21\xspace}
\newcommand{\expertNo}{12\xspace}
\newcommand{\safetyNo}{6\xspace}
\newcommand{\isafetyNo}{36\xspace}
\newcommand{\iriskNo}{\edit{14}\xspace}
\newcommand{\strategyNoNum}{6\xspace}
\newcommand{\strategyNo}{six\xspace}

\newcommand{\eg}{e.g., }
\newcommand{\etc}{etc.}
\newcommand{\ie}{i.e., }
\newcommand{\etal}{et al.\@\xspace}

\renewcommand{\quote}{\list{}{\rightmargin=\leftmargin\topsep=4pt}\item\relax}

\newcounter{ctr}
\newcounter{savectr}
\newcounter{ectr}

\newenvironment{newitemize}{%
\begin{list}{\mbox{}\hspace{5pt}$\bullet$\hfill}{\labelwidth=15pt%
\labelsep=4pt \leftmargin=12pt \topsep=3pt%
\setlength{\listparindent}{\saveparindent}%
\setlength{\parsep}{\saveparskip}%
\setlength{\itemsep}{3pt} }}{\end{list}}


\newenvironment{newenum}{%
\begin{list}{{\rm (\arabic{ctr})}\hfill}{\usecounter{ctr} \labelwidth=17pt%
\labelsep=3pt \leftmargin=21pt \topsep=3pt%
\setlength{\listparindent}{\saveparindent}%
\setlength{\parsep}{\saveparskip}%
\setlength{\itemsep}{-1pt} }}{\end{list}}

\newlength{\saveparindent}
\setlength{\saveparindent}{\parindent}
\newlength{\saveparskip}
\setlength{\saveparskip}{\parskip}

\titlespacing\section{0pt}{12pt plus 4pt minus 2pt}{3pt plus 2pt minus 2pt}
\titlespacing\subsection{0pt}{12pt plus 4pt minus 2pt}{3pt plus 2pt minus 2pt}

\newcommand{\titheading}[1]{{\vspace{5pt}\noindent\sc{#1}}}
\newcolumntype{L}{>{\raggedright\arraybackslash}p} 
\newcolumntype{C}{>{\centering\arraybackslash}p} 
\newcolumntype{R}{>{\raggedleft\arraybackslash}p} 

\newcounter{mytable}
\def\mytable{\begin{centering}\refstepcounter{mytable}}
\def\endmytable{\end{centering}}

\def\mytablecaption#1{\vspace{2mm}
                      \centerline{Table \arabic{mytable}.~{#1}}
                      \vspace{6mm}
             \addcontentsline{lot}{table}{\protect\numberline{\arabic{mytable}}~{#1}}}

\newcounter{myfig}
\def\myfig{\begin{centering}\refstepcounter{myfig}}
\def\endmyfig{\end{centering}}

\def\myfigcaption#1{
             \vspace{2mm}
             \centerline{\textsf{Figure \arabic{myfig}.~{#1}}}
             \vspace{6mm}
             \addcontentsline{lof}{figure}{\protect\numberline{\arabic{myfig}}~{#1}}}
\begin{abstract}
Research involving at-risk users---that is, users who are more likely to experience a digital attack or to be disproportionately affected when harm from such an attack occurs---can pose significant safety challenges to \edit{both} users and researchers.
Nevertheless, \edit{pursuing research in computer security \& privacy (S\&P) is crucial to understanding how to meet the digital-safety needs of at-risk users and to design safer technology for all.}
To standardize and bolster safer research involving such users, we offer an analysis of \paperNo academic works to elicit \iriskNo research risks and \isafetyNo safety practices used by \edit{a} growing community of researchers. 
We pair this inconsistent set of reported safety practices with oral histories from \expertNo domain experts to contribute scaffolded and consolidated pragmatic guidance that researchers can use to plan, execute, and share safer digital-safety research involving at-risk users.
We conclude by suggesting areas for future research regarding the reporting, study, and funding of at-risk user research.
\end{abstract}
\section{Introduction}
\label{sec:intro}
\noindent A growing body of research in computer
security \& privacy (S\&P) and human-computer interaction (HCI) is drawing attention to the digital security, privacy, and safety (i.e., \textit{digital safety}) needs and experiences of
at-risk users~\cite{warford2022atrisk, sannon_privacy_2022,
bhalerao_ethical_2022}.
These works cover users facing a spectrum of risks, including those who face an immediate threat of experiencing a digital attack
(survivors of intimate partner violence (IPV)~\cite{freed_stalkers_2018},
political activists~\cite{daffalla_defensive_2021}); an
increased likelihood to be targeted (LGBTQIA+
people~\cite{devito_values_2021}, political
campaigners~\cite{consolvo_why_2021}); or disproportionate harm
from an attack (children~\cite{piccolo_chatbots_2021}, people
experiencing homelessness~\cite{sleeper_tough_2019}).

It is crucial to understand and address the digital-safety needs and experiences of at-risk users, not only to develop effective mitigation approaches, but because improving their digital safety can benefit everyone.
Nevertheless, conducting research that involves at-risk users can be daunting.
When studying people who may be under a heightened threat of surveillance, harassment, or other digitally-mediated attacks, standard research activities---like recruiting, scheduling, providing participation incentives, and reporting results---may exacerbate risk to research participants and the group(s) they represent, as well as introduce risk to the researchers conducting the work.
Thus any research involving at-risk users requires extra caution in order to ensure safety, which can slow research progress.

In this paper, we systematize knowledge from the S\&P and HCI research communities to develop pragmatic guidance about reducing risk of harm in the planning, execution, and sharing of digital-safety research involving at-risk users (i.e., \textit{at-risk research} hereafter).
Our guidance reflects a systemization of ``good'' practices based on an analysis of \paperNo academic works and oral histories from an expert panel of S\&P scholars, as guided by the following:

\vspace{5pt}
\begin{enumerate}[leftmargin=1cm,label=\bfseries Q\arabic*:]
    \item What digital-safety risks are associated with research  involving at-risk users?
    \item What practices do researchers report employing to help mitigate digital-safety risk in at-risk research?
    \item What pragmatic guidance might researchers follow to reduce the risk of harm in their digital-safety research involving at-risk users?
\end{enumerate}
\vspace{5pt}

From our analysis of the \paperNo academic works, we identified \isafetyNo safety practices researchers reported using to address \iriskNo explicitly articulated risks.
We found a wide variety of practices, but only sparse and inconsistent reporting of them: the vast majority of works lacked sufficient detail to assess the safety of research procedures, and, by extension, to replicate safety practices.
Furthermore, there were no widely recognized standards for reporting on digital-safety practices.
The lack of consistency in reporting safety practices suggests that strategies for reducing the risks inherent with at-risk research may not be widely known or accepted.

To develop pragmatic guidance, we engaged S\&P scholars \edit{(all co-authors)} who have substantive experience in at-risk research.
Building on written and oral histories \cite{NAS_2009, campbell_its_2016}, we formulated \strategyNo strategies for safer digital-safety research involving at-risk users.
Such strategies propose that researchers assess and mitigate risks via threat modeling, \edit{select the lowest risk method that addresses the research goals}, and handle data and publication with care, among others.
The strategies complement well-documented guidance on more general ethical research approaches~\cite{munteanu_situational_2015, gustafson_ethics_2014, protections_ohrp_belmont_2010} by adding a strong focus on identifying and mitigating risk introduced by digital-safety research involving at-risk users.
We hope this \edit{systematization of knowledge} will help the community work toward more consistent \edit{use and} reporting of safety strategies for methods used in at-risk research.
\section{Background \& Related work}
\label{sec:related-intro}
\noindent In this section, we define key terms, then review methodological approaches to at-risk research from the S\&P and HCI communities. 

\label{sec:related-defining}
\paragraph{Digital-safety research.}
We use \textit{digital-safety research} to refer to research about a person's or a group's state of security, privacy, safety, and autonomy, as it relates to their digital footprint.
While there are multiple definitions of \textit{safety} \cite{antle_ethics_2017, goerzen2019entanglements}, we use it to mean when technologies serve, enable, or empower activities rather than being sources of harm (e.g., vectors for harassment, surveillance).

\paragraph{Risk, harm, \& research harm.}
\textit{Risk} is the \edit{probability that} a person---independently or as part of a group---will experience danger or harm \cite{luhmann_risk_2017}. 
\textit{Harm} is a negative impact to a person's psychological/emotional, physical, financial/economic, or social/relational condition \cite{field_glossary_2004, scheuerman2021framework}, including injury to safety, rights, or welfare.
Harm has been characterized across three dimensions---probability, severity, and duration \cite{hebert_fulfilling_2015}; severity has multiple dimensions, such as intent, scale, and urgency~\cite{scheuerman2021framework}.
We use \textit{research harm} to refer to harm caused to participants, the group(s) they represent, the public, or researchers themselves as a result of research activities \cite{protections_ohrp_belmont_2010}.
For instance, negligent research approaches may destabilize communities on- and offline (e.g., via deanonymization) \cite{chancellor_taxonomy_2019, leal_activism_2021}, and research prototypes can lead to harmful, unintended consequences.
In this paper, we focus on harm resulting from the planning, execution, or sharing of research involving at-risk users. 

\paragraph{At-risk users.}
\label{sec:related-groups}
We define a user(s) as being \textit{at-risk} if they face an elevated likelihood of an attack to their digital safety, have factors that influence or exacerbate their chances of being targeted, and/or experience heightened harm as a result of a digitally-mediated attack \cite{warford2022atrisk}.
Influential factors may be due to \textit{societal factors} (e.g., politics, marginalization), \textit{relationship factors} (e.g., relying on a third party for digital support, having a relationship with an attacker) and \textit{personal circumstances} (e.g., prominence in comparison to others, having access to a sensitive resource) \cite{warford2022atrisk}.
Identifying differences in risk is crucial to not flatten at-risk users \edit{under} one universal banner, which could result in inadequate harm mitigation \cite{warford2022atrisk, sannon_privacy_2022}.

Recent scholarship in S\&P and HCI has covered various topics related to at-risk users, such as the digital-safety needs of political activists~\cite{alvarado2017making,daffalla_defensive_2021,kow_mediating_2016,marczak_social_2017}, survivors of IPV \cite{freed_digital_2017,chatterjee2018spyware,freed_is_2019,havron2019clinical,matthews_stories_2017}, people experiencing homelessness~\cite{sleeper_tough_2019} or incarceration~\cite{owens_you_2021}, and people who experience identity-based marginalization within society (e.g., queer and trans people~\cite{ geeng_like_nodate}, women \cite{im2021yes,sultana2018design}, and people of color~\cite{boyd_understanding_2021}). 
From this growing volume of empirical research, \edit{works have synthesized contextual risk factors \cite{warford2022atrisk} and privacy risks for marginalized groups \cite{sannon_privacy_2022} but few directly address how to do digital-safety research safely.}


Scholars have developed methodological guidance for research with specific at-risk groups and surveyed some research approaches to representing and evaluating risk and security \cite{goerzen2019entanglements, scheuerman2021framework}. 
Slupska et al. \cite{slupska2021participatory} propose participatory threat modeling to elicit often overlooked risk factors of a single at-risk user group (survivors of IPV) for technology design, while 
Bhalerao et al. \cite{bhalerao_ethical_2022} call for work with at-risk users to employ confidentiality practices, be attuned to experiences of trauma, and adopt justice-oriented principles at specific (but not all) stages of research. 
In their review of marginalized users, Sannon and Forte~\cite{sannon_privacy_2022} ask that researchers develop shared best practices, including suggestions for fairer participant reimbursement and author position statements.
Collectively, these works reveal a need for theoretical and methodological convergence, to draw together HCI expertise in treating at-risk users sensitively with S\&P expertise in mitigating digital-safety harm.

\paragraph{Digital-safety \& ethical practices.}
Within the S\&P and HCI research communities, it is generally considered common knowledge that research might increase harm to at-risk users. 
However, these communities lack agreed-upon best practices for measuring, reporting, and implementing approaches to combat the adverse effects of researcher involvement or published works.
Some research ethics guidelines and practices have been established to limit or minimize risk of adverse or harmful outcomes in human subjects research broadly.
For example, researchers must often secure approval from ethical review committees (e.g., institutional review boards) before conducting human subjects research. 
Many program committees and journal editors have also established ethical requirements---with varying formality---for publications in their venues; they may reject or retract submissions on ethical grounds (e.g., \cite{sandp_ieee_nodate}). 

For human subjects research in general, the S\&P and HCI communities have institutionalized 
many ethical practices, such as obtaining informed consent from research participants~\cite{kenneally_menlo_2012, protections_ohrp_belmont_2010}.
Similarly, the S\&P community has longstanding norms about when and how to disclose new security vulnerabilities and attacks with the aim of limiting harm \cite{sandp_ieee_nodate}. 
These practices and norms are critically important for promoting ethical research, but they are not necessarily sufficient for research involving at-risk users.
As an example, research ethics discussions rarely encourage digital safety-enhancing behaviors by the research team when conducting indirect (i.e., no direct interaction) research about at-risk users \cite{pater_no_2022} or enforce the use of practical safety plans \cite{bellini_fragments_2021}.
While all digital-safety practices could be considered forms of ethical practice, not all ethical practices concern safety (e.g., conversations around standards of conduct and moral values).
Adopting a safety-conscious approach helps raise awareness of the importance of safety as a specific area of focus in research ethics.

\paragraph{Open questions.}
While the aforementioned prior works are valuable for \ds research involving at-risk users, they have several limitations.
Although Sannon and Forte reviewed recent privacy research about marginalized users, some at-risk users are not considered marginalized (e.g., journalists) and experience digital-safety risks beyond threats to privacy, including threats to their security or safety.
Frameworks or meta-analyses that address the digital safety of at-risk users (e.g., \cite{warford2022atrisk,goerzen2019entanglements, baruh_online_2017, thomas_sok_2021}) do not discuss how research itself may further harms.
Although most reviews~\cite{bhalerao_ethical_2022, warford2022atrisk, thomas_sok_2021} conclude with calls for clarity on best practice guidelines or recommendations for performing research safely, no prior works---to our knowledge---address the entire research process (as opposed to specific research stages \cite{bhalerao_ethical_2022,sannon_privacy_2022, antle_ethics_2017})---despite calls to do so \cite{thomas_sok_2021, warford2022atrisk, baruh_online_2017,goerzen2019entanglements,bhalerao_ethical_2022,sannon_privacy_2022,antle_ethics_2017}.
\section{Methods: Analysis of existing works}
\label{sec:methods-litrev}
\noindent To systematize existing approaches to safety in at-risk research, we start with an analysis of \paperNo academic works. 
We use the results of this analysis (presented in \secref{sec:review}) and knowledge from eight experienced S\&P scholars to generate pragmatic strategies to guide safer digital-safety research involving at-risk users (presented in \secref{sec:six-strategies}).
Four authors (first, second, third, and fifth) were involved in the analysis of existing works (the \textit{analysis team}), while all \expertNo authors (referred to as an \textit{expert panel}) contributed to the systematization of practices into strategies.

The analysis of existing works followed a rapid evidence review~(RER) methodology~\cite{varker_rapid_2015,moons_rapid_2021}. 
RERs are a rigorous approach that provide relevant and actionable evidence to strengthen policy and practice. 
They follow the same steps as a systematic review for identifying, selecting, critically evaluating and analyzing data, but some components are simplified (i.e., exclusion of grey literature) to ensure results are delivered quickly.
In doing so, we followed established guidelines for conducting a research synthesis via an RER \cite{moons_rapid_2021}, as described below.

\paragraph{Corpus.}
\label{sec:related-search}
To understand how research practices with at-risk users are being reported, we analyzed \paperNo peer-reviewed papers.
We focused our analysis on digital-safety research (rather than technology research more broadly), because we expected that such works would be more likely to report their digital-safety considerations.

To begin, we retrieved an existing open dataset of 127 digital-safety papers from Warford et al. \cite{warford2022atrisk} \edit{(which include several authors of this work)} who analysed 31 distinct at-risk population categories (e.g., journalists, sex workers).
We used specialist databases including USENIX, IEEE Xplore, and the ACM Digital Library to identify additional works that were published in premier S\&P and HCI venues after this initial dataset was collected: CCS, CHI, CSCW, IEEE S\&P, NDSS, PETS, SOUPS, and USENIX Security.
Using the same inclusion criteria\footnote{
To be included in the Warford et al. \cite{warford2022atrisk}, dataset, the \textit{primary} purpose of the work had to explore: (a) at-risk users \textit{and} (b) digital-safety threats.
Papers had to be full-length, peer-reviewed, and written in English (6,428 works). 
127 papers were identified through a rigorous approach of determining relevancy; \edit{the dataset for this work} is available at \bhref{https://docs.google.com/spreadsheets/u/1/d/e/2PACX-1vTy06wCGf8h4nKqU_2EB7E5b-gXKQGz0uOB6JAwHGjzBaurTbLRxEc0AcXOMIH_h5OxLTHM5-4jzg8A/pubhtml}{this link}. }, 
we considered  \edit{3,948} papers published between late 2020 and  202\edit{2}. 
Only \edit{65} of these works satisfied the criteria for inclusion.
Four papers in our corpus each referenced a prior work containing further details about safety procedures. 
We therefore  included those four additional papers, resulting in a total of \paperNo papers.\footnote{The complete list of \paperNo papers can be found at \bhref{https://docs.google.com/spreadsheets/d/e/2PACX-1vSjMiPhJyHyZPF0NUQFfEoB3yEUTZlr9z1SKmAbCE0T0_lTwqENBXmOXvon68eKdhG_WBYIT25YMWR2/pubhtml?gid=1991494022&single=true}{this link}.}

\paragraph{Data extraction criteria and coding process.}
\label{sec:method-data-extraction}
To develop a set of data extraction criteria, the first author read a randomly selected subset of \edit{20} papers of the corpus (\edit{9.8\%}) to identify exemplar and outlier cases.
The third author reviewed the coding \edit{criteria} to ensure a broad coverage of risks and safety practices.
Once this coding \edit{criteria} was reviewed by two other authors, the first author then read each of the \paperNo works, and coded the reported methodological approaches to at-risk research (i.e., stated methods), ethical dimensions, and any safety practices performed by the works' authors.
The coding process of the \paperNo papers was performed using an established protocol by one experienced coder, and results checked by a second author (similar to most RER approaches \cite{moons_rapid_2021}).
The results synthesis and clarification of evidence (described next) was performed by the first, second, and fifth authors in iterative steps \cite{varker_rapid_2015}.

\paragraph{Data analysis.}
\label{sec:method-data-analysis}
To answer \textbf{Q1}, the analysis team developed, discussed, and refined our findings into \iriskNo common risks that authors of the works reported as relevant to their research methods (see \tabref{tab:risks} in \secref{sec:review}).
To answer \textbf{Q2}, the process was repeated to identify \isafetyNo common, \edit{distinct} safety practices; that is, strategies and actions that authors of the works reported using to support the safety of participants, the at-risk users they represented (i.e., non-participants), or research team members (see \tabref{tab:sok-ds} in \secref{sec:safety-practices}).
The analysis team considered a practice \textit{common} if it was reported more than three times among (a) research with the same at-risk group, or (b) research with at least three different at-risk groups.
Open codes were then synthesized into two abstract categories (independent of practice employed or at-risk user group involved): \textit{research risks} and \textit{digital-safety practices}. 

\paragraph{Further systematization with S\&P scholars.}
\label{sec:method-sok-creation}
In aggregate, existing works provided significant information about risks and safety practices involved in at-risk research, but did not provide pragmatic strategies for guiding such research. 
To answer \textbf{Q3} and to complement our RER, we engaged eight S\&P scholars to elicit further knowledge on research practices and help formulate actionable safety strategies. 
These scholars, who may be considered as domain experts, are authors on this paper to represent their contributions to this work. 
We discuss the process used to develop the strategies in \secref{sec:six-strategies}.

Expert engagement is known to increase the relevance, use, and dissemination of RERs~\cite{varker_rapid_2015, moons_rapid_2021}, and we hypothesized that expert accounts would provide invaluable information on digital safety that was absent from many works---such oral histories and traditions are often used to teach safe and ethical practices~\cite{NAS_2009}.

\paragraph{Limitations.}
Despite our approach, we cannot overcome bias \cite{moons_rapid_2021, varker_rapid_2015} that occurs when authors, reviewers, or publishers favor specific studies (e.g., promoting methodological orientations, studies of large sample sizes, etc.).
Reviewers may request omissions of methodological information (which we have experienced \edit{ourselves in some of our prior work}), or the authors may omit such information.
We do not presume that absence of methodological descriptions on digital safety equates to unsafe research, but rather that such authors were either not encouraged or did not feel compelled to report their safety practices.
Thus, our RER cannot draw conclusions about how safety practices were enacted, but we can comment on how such practices were accounted for, described, and justified \edit{in publications}.

Our data was gathered systematically and is large enough to cover common digital-safety challenges in at-risk research as performed by the S\&P and HCI communities. 
\edit{As a first systemization on this topic, we scoped our efforts to the research community’s knowledge and felt it was appropriate to focus on a representative set of academic works by the privacy and security community.}
\edit{The} works we analyzed were written in English and took place in predominantly Western contexts, most often in the U.S.
Consequently, this systematization may skew toward Western audiences, which may not accommodate all digital-safety considerations in at-risk research.
\begin{table*}
\scriptsize
\centering
\resizebox{2\columnwidth}{!}{%
\renewcommand{\arraystretch}{1.3}
\begin{tabular}[t]{@{}p{0.75in}p{1in}lp{3in}l@{}}
\toprule
\CC{40} \textbf{Risks posed} & \CC{40}  & \CC{40}  & \CC{40} \textbf{Description} & \CC{40}  \textbf{Example papers} \\
\midrule
to participants
  & \ldots from data collection
  & \CC{15}Breach of confidentiality
    & \CC{15}Researchers may be compelled to disclose \edit{participant data} to an authority without participants' consent, due to subpoena, duties to law enforcement, or parental rights.
    & \CC{15}\cite{wisniewski_parents_2017, boyd_understanding_2021, havron2019clinical, heyer_opportunities_2020} \\
  & & Unauthorized access
      & Even when using best-practice data-security tools, \edit{adversaries may gain unauthorized access to sensitive participant data}.
      & \cite{mcdonald_its_2021, mcgregor_investigating_2015}
      \\
\cmidrule(l){2-5}

 & \multirow{3}{*}{\parbox{1in}{\raggedright\ldots from direct research, including primary interviews or when researchers offer digital-safety advice}}
 
  & \CC{15}Coerc\edit{ion of contributions}
      & \CC{15}Adversaries may accompany participants to studies and provide or discourage responses, especially when the adversary is an intimate (e.g., a partner, family member, or caregiver).
      & \CC{15}\cite{havron2019clinical, mentis_upside_2019, freed_stalkers_2018, mchugh_most_2017}
      \\  
  & & \edit{Disruption to support}
      & \edit{Researchers may disrupt the normal functioning of digital-safety services and place a participant's security in jeopardy. Participants may also conflate research activities with service provision} and feel compelled to participate in research to receive support.
      & \cite{bhalerao_ethical_2022,freed_is_2019}
      \\
  & & \CC{15}\edit{Distress and} re-traumatization
      & \CC{15} At-risk participants may be prompted to recount moments where they experienced digital-safety harms, which may cause distress. \edit{This can extend to viewing the researcher as a physical threat to a participant's wellbeing.}
      & \CC{15}\cite{andalibi_lgbtq_2022, freed_stalkers_2018, havron2019clinical, to_they_2020, chen_trauma-informed_2022}
      \\
  & & Escalation of abuse
      & Research activities may require \edit{or encourage} participants to break routines or take protective actions like removing spyware, which may incite adversaries to escalate their abuse \edit{or retaliate against the participant.} 
      & \cite{marczak_social_2017,havron2019clinical,mcgregor_investigating_2015,tseng_care_2022}
      \\
  & & \CC{15}\edit{Withhold benefit}
      & \CC{15}If researchers do not inform participants \edit{about the viability of reported threats or available protective practices}, participants may be at greater risk.
      & \CC{15}\cite{sambasivan_privacy_2018,lastdrager_how_2017}
      \\
\cmidrule(l){2-5}

 & \multirow{3}{*}{\parbox{1in}{\raggedright\ldots from the publication of research products}}
  & Adversarial feedback
      & Research may publicize protective strategies in ways that inform adversaries, who then correspondingly \edit{adapt or} escalate their attacks.
      & \cite{doerfler_im_2021, bellini_so-called_2021, freed_stalkers_2018, tseng_tools_2020, boyd_understanding_2021, matthews_stories_2017}
      \\
  & & \CC{15}Deanonymization
      & \CC{15} Unsuccessfully paraphrased quotes \edit{or poor redaction of participant information} might reveal the identities of at-risk participants, particularly those who are public figures.
      & \CC{15}\cite{consolvo_why_2021, freed_stalkers_2018, freed_digital_2017}
      \\
  & & Misrepresentation
    & Research may inadvertently mischaracterize participants' digital-safety needs, which may disrupt their safety strategies or encourage risky or ineffective interventions.
    & \cite{mcdonald_its_2021,scheuerman_safe_2018,mentis_upside_2019}
      \\
\midrule

 to researchers &
  & \CC{15}Burnout and vicarious trauma
      & \CC{15}Immersion in stories of hate, harassment, and abuse may incur vicarious trauma or secondhand traumatic stress, which may result in burnout or exhaustion.
      & \CC{15}\cite{chen_trauma-informed_2022, moitra_parsing_2021, nova_facebook_2021, freed_is_2019, tseng_digital_2021, andalibi_understanding_2016}
      \\
  & & Harassment \edit{and intimidation}
     & Researchers may themselves experience hate and harassment due to public statements about their research. Scholars with marginalized identities are particularly susceptible.
     & \cite{doerfler_im_2021, andalibi_lgbtq_2022}
     \\
  & & \CC{15}\edit{Liability exposure}
      & \CC{15}\edit{Researchers may be subject to criminal prosecution or civil litigation for failing to disclose observed vulnerabilities \edit{(of at-risk groups or technical systems)} uncovered during their research.}
      & \CC{15}\edit{\cite{mchugh_most_2017, venkatasubramanian_exploring_2021, boyd_understanding_2021}}
      \\
  & & \edit{Surveillance}
    & \edit{Adversaries \edit{who} have strategies for \edit{digitally} tracking and monitoring at-risk groups may extend \edit{these tactics} to researchers.}
    & \edit{\cite{shklovski_use_2018, owens_you_2021, sanches_under_2020}}
    \\
\bottomrule
\end{tabular}%
}
\vspace*{2mm}
\caption{Risks posed by digital-safety research involving at-risk users, systematized from our analysis.}
\label{tab:risks}
\vspace{-2.5mm}
\end{table*}

\section{Reported risks \& Safety practices}
\label{sec:review}
\noindent \edit{From these \paperNo works, we identified \iriskNo commonly anticipated risks incurred by digital-safety research to participants, the group(s) they represented, and researchers (Table \ref{tab:risks}), as well as the practices researchers used to mitigate those risks (\tabref{tab:sok-ds}).
This compilation of risks is non-exhaustive as they are products of authors', reviewers', and venues' publication norms, but can provide a useful baseline for digital-safety research.
We delineate these risks via \textit{person affected} (participants and  at-risk users, researchers) and by \textit{research processes} (data collection, direct research encounters, publication of  deliverables), which are illustrated by examples of \textit{specific risks} (e.g., escalation of abuse).}


\subsection{Risks to participants and at-risk users}
\label{sec:risks-participants}
\noindent \edit{Research involves risks for all participants \cite{kenneally_menlo_2012}, but at-risk users 
can face greater dangers from data collection, direct research encounters, and publications.} These risks apply most directly to participants themselves, but can also extend to other, non-participating members of the groups the participants represent.

\paragraph{From data collection.} \edit{Empirically grounded research into at-risk users' digital safety involves collecting data about them, typically in a manner designed to minimize their discomfort and enhance their wellbeing.}
\edit{When either indirect methods (e.g., scraping data) or direct interaction are used to elicit at-risk user data, the risk of \textit{unauthorized data access} increases. 
Parties gaining unauthorized access may be relatively benign, such as colleagues outside the research team, or actively malicious, such as nation-states.
Adversaries may gain access to sensitive material or leak participants' research contributions to other harmful parties (e.g., journalists under surveillance \cite{mcgregor_investigating_2015}).
Interception of recruitment materials also qualifies, as it could identify membership in a group of at-risk users (e.g., sex workers seeking anonymity~\cite{mcdonald_its_2021}). }

\edit{For well-resourced adversaries, a breach of research data may occur via a \emph{compelled disclosure}, where researchers could be forced to divulge data to third parties without participants' consent.
For example, an adversary might use a subpoena from a legal authority (or even the threat of one) to compel a researcher to disclose what their participants told them \cite{heyer_opportunities_2020, havron2019clinical} or destroy information collected about an adversary.
If participants disclose their participation in illegal activities~\cite{boyd_understanding_2021}, or the researcher identifies elder or child abuse~\cite{wisniewski_parents_2017}, the researcher may be required to notify relevant authorities.
These disclosures can exacerbate security threats for some at-risk users, like activists \cite{boyd_understanding_2021} or sex workers \cite{mcdonald_its_2021}, who may take steps to avoid law enforcement.}

\paragraph{From direct research encounters.}
\edit{Direct research encounters with at-risk participants could also create \edit{undue} risk.
Users at risk of oppression or stigmatization \cite{warford2022atrisk} may face coercion by adversaries, inhibiting them from freely participating in research.
Data collected may be compromised if researchers do not consider the potential for participants to 
self-censor if, e.g., they are intimidated by adversaries or consider the research site unsafe~\cite{alghamdi_security_2015}. 
Participants may even be compelled to attend research encounters with adversaries such as partners~\cite{freed_digital_2017, freed_stalkers_2018}, family members~\cite{mchugh_most_2017}, or caregivers~\cite{mentis_upside_2019}.}

\edit{Participants who disclose information that adversaries disapprove of may face retaliation, harming their job prospects \cite{sannon_toward_2022, slupska_they_nodate}, access to resources \cite{tseng_care_2022}, or physical safety \cite{tseng_digital_2021}.
Similarly, participation in research that  requires a user to change their protective actions, such as locking an adversary out of a compromised account~\cite{havron2019clinical, tseng_care_2022, mcgregor_investigating_2015}, could incur an \emph{escalation of abuse}. 
Such situations are highly dangerous, creating an acute risk of other harms, such as physical stalking or monitoring~\cite{marczak_social_2017}.}


\edit{Digital-safety research may also \emph{distress} or \emph{re-traumatize} at-risk participants, by asking them to recount some of their most sensitive experiences \cite{andalibi_lgbtq_2022, freed_stalkers_2018}, or inadvertently \edit{triggering} feelings of judgment or shame that they did not take ``better steps'' to protect themselves from their adversaries \cite{to_they_2020, havron2019clinical} (c.f.,~\cite{chen_trauma-informed_2022}).
Researchers may wish to help by sharing guidance on S\&P best practices, but this must be done carefully, as improperly advising participants \edit{can} \emph{withhold benefit} from them or even cause additional harm. 
This could include omitting relevant advice about protective practices \cite{sambasivan_privacy_2018} or underplaying the severity of potential threats, leading \edit{participants} to make unsafe choices about securing their safety \cite{lastdrager_how_2017}.
Even well-informed advice, however, \edit{runs} the risk of disrupting dedicated S\&P support. 
For instance, participants may feel compelled to take part in research activities to get much-needed help \cite{freed_is_2019, bhalerao_ethical_2022}.}


\paragraph{From publication of research deliverables.}
\edit{Research deliverables like papers and reports may pose risks to at-risk groups downstream of direct research encounters. \emph{Misrepresentation} of participants' experiences of digital risk and harm could perpetuate myths about the causes of their vulnerability~\cite{mcdonald_its_2021} or enforce unfair and negative stereotypes~\cite{scheuerman_safe_2018}, as well as potentially leading people in power to create ineffective interventions~\cite{mentis_upside_2019}.
Further, as many at-risk groups depend on confidentiality and anonymity as a protective practice, the risk that a reader may \emph{de-anonymize} them in reports or papers could have severe consequences.
Researchers may include idosyncratic details about participants' experiences \cite{consolvo_why_2021}, add demographics that facilitate jigsaw identification \cite{freed_is_2019}, or advertise the location of private communities which may be subsequently targeted.}

\edit{S\&P publications may motivate readers to eliminate barriers for at-risk groups \cite{warford2022atrisk}, but may also be read by adversaries who then target at-risk users or incite others to do so ~\cite{doerfler_im_2021}.
These risks include providing instructions about how to attack a particular group of at-risk users, or naming specific software tools or online communities where adversaries can find further information~\cite{tseng_tools_2020, freed_stalkers_2018, bellini_so-called_2021}. 
Even activities that may initially seem beneficial could prove harmful to at-risk groups. For example, compiling disparate useful resources into reviews or best-practice guides could inadvertently reduce the burden of threat intelligence for adversaries \cite{boyd_understanding_2021, matthews_stories_2017, tseng_tools_2020}.}

\subsection{Risks to researchers}
\label{sec:risks-researchers}
\noindent \edit{Researchers can also incur emotional, physical, and legal harms that are under-reported in research deliverables \cite{moncur_emotional_2013}.}

Scholars can face \emph{vicarious trauma} from repeated exposure to dark and distressing content like stories of digital-safety harms. 
Vicarious trauma, also known as `second-hand' traumatic stress, is the emotional residue of exposure to traumatic stories and experiences \cite{chen_trauma-informed_2022, moitra_parsing_2021}.
Research activities---like witnessing active discrimination against a group of at-risk users by other members of society~\cite{nova_facebook_2021}, hearing accounts of technology-facilitated abuse~\cite{freed_stalkers_2018, freed_is_2019,tseng_digital_2021}, or reading online accounts of sexual violence in community groups~\cite{andalibi_understanding_2016}---can all evoke vicarious trauma.
As digital-safety researchers must conduct these activities in the course of doing their jobs, this research puts them at risk of \emph{burnout}, or exhaustion from work-related stress \cite{moitra_parsing_2021}.
\edit{Experience of exhaustion may be exacerbated for researchers who are \textit{exposed to legal liability} challenges.
Researchers may face civil or criminal charges if appropriate parties or companies are not informed of relevant information \cite{sun_child_2021}, such as abuse of persons or systems \cite{bellini_fragments_2021, bellini_choice-point_2020}.}

Researchers may \edit{become the target of} \emph{harassment and intimidation} in the process of publicizing their research findings, especially when their work revolves around ideas and concepts that may cause backlash, like the digital-safety needs of persecuted users \cite{doerfler_im_2021}. 
Scholars with marginalized identities are known to be particularly vulnerable \cite{andalibi_lgbtq_2022, doerfler_im_2021}.
\edit{Such actions may also extend to \emph{surveillance} of researchers by adversaries who may take an interest in pursuing at-risk groups, thereby violating the privacy and physical safety of researchers involved in conducting this work.}
\subsection{Safety practices}
\label{sec:safety-practices}
\begin{table*}[ht]
\centering
\resizebox{2\columnwidth}{!}{%
\begin{tabular}{@{}p{1.3in}p{0.15in}p{3.8in}p{2.2in}@{}}
\toprule
\CC{40} 
\textbf{Category} & \CC{40}
  \textbf{ID} & \CC{40}
  \textbf{Digital-safety practices} & \CC{40}
  \textbf{\edit{Example} papers} 
\\ \midrule
\multirow{4}{*}{\begin{tabular}[c]{@{}l@{}}Professional partnerships \\ \& Ethical review\end{tabular}} &
  SP1 & 
  Elicit expert (academic) opinion on topic area &
  \cite{sambasivan_they_2019, matthews_stories_2017, barwulor_disadvantaged_2021, mcdonald_its_2021, chen_trauma-informed_2022, jicol_designing_2022, kim_personalization_2022, sultana_shishushurokkha_2022, thomas_its_2022} \\
 &
  \CC{15} SP2 & \CC{15}
  Form professional partnerships (e.g., support services for at-risk users) & \CC{15}
  \cite{marczak_social_2017, kozubaev_spaces_2019, freed_stalkers_2018, vitak_i_2018, guberek_keeping_2018, sleeper_tough_2019, matthews_stories_2017, nicholson_training_2021, page_perceiving_2022, slupska_they_nodate} \\ 
 & 
  SP3 &
  Invite and include an at-risk user to join research team &
  \cite{sambasivan_they_2019, barwulor_disadvantaged_2021, mcdonald_its_2021, musgrave_experiences_2022} \\
 &
  \CC{15} SP4 & \CC{15}
  Seek external (non-institutional) ethical review approval or monitoring & \CC{15}
  \cite{freed_stalkers_2018, freed_is_2019, chen_computer_2019, li_there_2021}
\\ \midrule

  \multirow{7}{*}{\begin{tabular}[c]{@{}l@{}}Positionality \& \\ Participant engagement \end{tabular}}
 &
  SP5 &
  Build rapport with participants for understanding digital-safety needs &
  \cite{to_they_2020, lastdrager_how_2017, sambasivan_privacy_2018, consolvo_why_2021, daffalla_defensive_2021, moitra_parsing_2021, musgrave_experiences_2022, adler_burnout_2022, coles-kemp_protecting_2022} \\
 &
  \CC{15} SP6 & \CC{15}
  Conduct pilot studies with general (non-at-risk) users & \CC{15}
  \cite{obada-obieh_towards_2020, akter_i_2020, chen_computer_2019, jicol_designing_2022, im_womens_2022, coles-kemp_protecting_2022, munyendo_desperate_2022} \\
 &
  SP7 &
  Conduct studies with proxies for at-risk users (e.g., advocacy groups) &
  \cite{owens_you_2021, kim_personalization_2022, latulipe_unofficial_2022, sultana_shishushurokkha_2022, bhalerao_ethics_2022, coles-kemp_protecting_2022, afnan_aunties_nodate} \\
 &
  \CC{15} SP8 & \CC{15}
  \edit{Include researchers whose identities affirm participants'} & \CC{15} 
  \cite{sambasivan_privacy_2018, sambasivan_they_2019, daffalla_defensive_2021, alghamdi_security_2015, rifat_purdah_2021, musgrave_experiences_2022, sultana_shishushurokkha_2022, im_womens_2022, afnan_aunties_nodate, slupska_they_nodate} \\
 &
  SP9 &
  Practice responsiveness in data collection sessions to potential threats &
  \cite{geeng_usable_2020, mehmood_towards_2019, starks_designing_2019, obada-obieh_towards_2020, sleeper_tough_2019, ahmed_privacy_2015, daffalla_defensive_2021, nova_facebook_2021, steinbrink_digital_2021, sultana_shishushurokkha_2022} \\
 &
  \CC{15} SP10 & \CC{15}
  Provide professional therapeutic support for emotive topics & \CC{15}
  \cite{sannon_privacy_2019, obada-obieh_towards_2020, andalibi_understanding_2016, chen_computer_2019, nova_facebook_2021, venkatasubramanian_exploring_2021, ali_understanding_2022, geeng_like_nodate, munyendo_desperate_2022} \\
 &
  SP11 &
  Train team members in working with digital-safety risks &
  \cite{shklovski_use_2018, sannon_privacy_2019, daffalla_defensive_2021, ali_understanding_2022} 
\\ \midrule
  \multirow{9}{*}{\begin{tabular}[c]{@{}l@{}}Privacy-preserving \\ data collection\end{tabular}}
  &
  \CC{15} SP12 & \CC{15}
  Discourage participant self-disclosure (e.g., personal histories) & \CC{15}
  \cite{le_blond_enforcing_2018, bowyer_understanding_2018, to_they_2020, simko_computer_2018, scheuerman_safe_2018, guberek_keeping_2018, venkatasubramanian_exploring_2021, ali_understanding_2022, kim_personalization_2022, adler_burnout_2022} \\
 &
  SP13 &
  Focus data collection on supporting participant safety needs &
  \cite{schmitt_participatory_2018, jensen_when_2020, shklovski_use_2018, simko_computer_2018, strohmayer_technologies_2019, masaki_exploring_2020, consolvo_why_2021, daffalla_defensive_2021, musgrave_experiences_2022, bhalerao_ethics_2022} \\
 &
  \CC{15} SP14 & \CC{15}
  Do not collect or ask for participant demographic data & \CC{15}
  \cite{schmitt_participatory_2018, mcdonald_realizing_2020, vitak_i_2018, sleeper_tough_2019, barwulor_disadvantaged_2021, owens_you_2021, boyd_understanding_2021, mcdonald_its_2021, thomas_its_2022, im_womens_2022} \\
 &
  SP15 &
  Do not collect personally identifiable information on participants &
  \cite{heyer_opportunities_2020, vashistha_threats_2019, freed_stalkers_2018, lastdrager_how_2017, guberek_keeping_2018, freed_is_2019, chen_computer_2019, mcgregor_investigating_2015, hamilton_risk_2022, munyendo_desperate_2022} \\
 &
  \CC{15} SP16 & \CC{15}
  \edit{Implement protocols for researchers to prevent stalking by adversaries} & \CC{15}
  \cite{marczak_social_2017, hornung_navigating_2017, chen_computer_2019} \\
 &
  SP17 &
  \edit{Separate} potential threats from at-risk users during data collection &
  \cite{kozubaev_spaces_2019, sannon_privacy_2019, alghamdi_security_2015, mchugh_most_2017, murthy_individually_2021, nova_facebook_2021, rifat_purdah_2021, musgrave_experiences_2022}\\
 &
  \CC{15} SP18 & \CC{15}
Permit participants to contribute false information (e.g., pseudonyms) & \CC{15}
  \cite{heyer_opportunities_2020, barwulor_disadvantaged_2021, nova_facebook_2021, li_there_2021, mcdonald_its_2021, hamilton_risk_2022} \\
 &
  SP19 & 
    Offer participants many modalities to contribute (e.g., audio, notes) & 
  \cite{petelka_being_2020, hayes_cooperative_2019, ahmed_addressing_2016, mentis_upside_2019, consolvo_why_2021, sultana_unmochon_2021, ali_understanding_2022, jicol_designing_2022, bhalerao_ethics_2022, sannon_privacy_2022} \\
 & 
  \CC{15} SP20 &
  Secure confidentiality and privacy of online and in-person research sites &
  \cite{freed_stalkers_2018, sambasivan_privacy_2018, freed_is_2019, lerner_privacy_2020, chen_computer_2019, alghamdi_security_2015, nova_facebook_2021, tseng_digital_2021, bhalerao_ethics_2022, slupska_they_nodate} 
\\ \midrule
  \multirow{4}{*}{\begin{tabular}[c]{@{}l@{}}Secure data storage \\ \& processing\end{tabular}} &
  SP21 &
  Implement strict data access control measures for research data &
  \cite{marczak_social_2017, sambasivan_they_2019, consolvo_why_2021, tseng_digital_2021, ali_understanding_2022, thomas_its_2022, wang_gay_2022, adler_burnout_2022, gruber_we_2022, slupska_they_nodate} \\
 &
  \CC{15} SP22 & \CC{15}
  Redact participant information prior to analysis by research team & \CC{15}
  \cite{vashistha_threats_2019, petelka_being_2020, sanches_under_2020, mcgregor_when_2017, steinbrink_digital_2021, sultana_unmochon_2021, zou_role_2021, tseng_care_2022, hong_analyzing_2022, munyendo_desperate_2022} \\
 & 
  SP23 &
  Use encryption for research data in-transit and at-rest &
  \cite{le_blond_enforcing_2018, guberek_keeping_2018, hornung_navigating_2017, obada-obieh_towards_2020, mcgregor_when_2017, mcgregor_investigating_2015, mcgregor_would_2017} \\
 &
  \CC{15} SP24 & \CC{15}
  Use non-encrypted safe storage for research data in-transit and at-rest & \CC{15}
  \cite{sanches_under_2020, mentis_upside_2019, consolvo_why_2021, chen_computer_2019, sultana_unmochon_2021, ali_understanding_2022, musgrave_experiences_2022, sultana_shishushurokkha_2022} 
\\ \midrule
  \multirow{5}{*}{\begin{tabular}[c]{@{}l@{}}Researcher accountability\end{tabular}} &
  SP25 &
  Conduct data collection sessions around participant schedules & 
  \cite{schmitt_participatory_2018, jensen_review_2005, cornejo_vulnerability_2016, rubya_interpretations_2017, steinbrink_digital_2021, tseng_digital_2021, musgrave_experiences_2022, hamilton_risk_2022, adler_burnout_2022} \\
 &
  \CC{15} SP26 & \CC{15}
  Offer formal proof of identity as professional researchers & \CC{15}
  \cite{sanches_under_2020, sannon_privacy_2019, sambasivan_they_2019, matthews_stories_2017, kim_personalization_2022, musgrave_experiences_2022} \\
 &
  SP27 &
  Only use data from publicly accessible sites (e.g., no authorization) &
  \cite{xue_right_2016, andalibi_understanding_2016, tseng_tools_2020, chung_privacy_2017, doerfler_im_2021, musgrave_experiences_2022, wang_gay_2022, owens_electronic_nodate} \\
 &
  \CC{15} SP28 & \CC{15}
  Provide proportional incentives to participants for contributions & \CC{15}
  \cite{kozubaev_spaces_2019, lastdrager_how_2017, vitak_i_2018, matthews_stories_2017, tseng_digital_2021, rifat_purdah_2021,  wilkinson_many_2022, im_womens_2022, hamilton_risk_2022, slupska_they_nodate} \\
 &
  SP29 &
  Be transparent with participants about risks incurred by research &
  \cite{khovanskaya_bottom-up_2020, hayes_cooperative_2019, sambasivan_privacy_2018, daffalla_defensive_2021, steinbrink_digital_2021, boyd_understanding_2021, rifat_purdah_2021, hamilton_risk_2022, bhalerao_ethics_2022, munyendo_desperate_2022} 
\\ \midrule
    \multirow{7}{*}{\begin{tabular}[c]{@{}l@{}}Sharing \& evaluating \\ deliverables \end{tabular}} &
   \CC{15} SP30 & \CC{15}
  Do not attribute reported data contributions with participant identifiers & \CC{15}
  \cite{mcdonald_realizing_2020, sanches_under_2020, consolvo_why_2021, hartikainen_safe_2021, ali_understanding_2022, alsoubai_friends_2022, sannon_privacy_2022, alomar_developers_2022, slupska_they_nodate} \\
 &
  SP31 &
  Do not report participant demographics in research deliverables &
  \cite{schmitt_participatory_2018, vitak_i_2018, freed_is_2019, lerner_privacy_2020, barwulor_disadvantaged_2021, li_there_2021, venkatasubramanian_exploring_2021, mcdonald_its_2021, bhalerao_ethics_2022, sannon_privacy_2022} \\
 &
  \CC{15} SP32 & \CC{15}
  Do not report participant names, pseudonyms, or identifiers & \CC{15}
  \cite{xue_right_2016, vashistha_threats_2019, shklovski_use_2018, kow_mediating_2016, vitak_i_2018, sanches_under_2020, obada-obieh_towards_2020, li_there_2021, alsoubai_friends_2022, geeng_like_nodate} \\
 & 
  SP33 &
  Paraphrase or withhold sources of data (e.g., websites they use) &
  \cite{xue_right_2016, khovanskaya_bottom-up_2020, simko_computer_2018, barwulor_disadvantaged_2021, doerfler_im_2021, li_there_2021, thomas_its_2022, alsoubai_friends_2022, afnan_aunties_nodate, hong_analyzing_2022} \\
 &
  \CC{15} SP34 & \CC{15}
  Evaluate research deliverables for adversarial feedback or education & \CC{15}
  \cite{freed_stalkers_2018, sambasivan_privacy_2018, matthews_stories_2017, consolvo_why_2021, daffalla_defensive_2021, hong_analyzing_2022} \\
 &
  SP35 &
  Selectively edit participant data in research deliverables &
  \cite{sleeper_tough_2019, andalibi_understanding_2016, hartikainen_safe_2021, doerfler_im_2021, tseng_digital_2021, ali_understanding_2022, tseng_care_2022, wilkinson_many_2022, alsoubai_friends_2022,wei_anti-privacy_nodate} \\
 & 
  \CC{15} SP36 & \CC{15}
  Provide participants control of their contributions (e.g., permit redaction) & \CC{15}
  \cite{le_blond_enforcing_2018, sanches_under_2020, gatehouse_troubling_2018, sambasivan_privacy_2018, moitra_parsing_2021, ali_understanding_2022, thomas_its_2022, hamilton_risk_2022, sannon_privacy_2022} 
\\ \bottomrule
\end{tabular}%
}
\caption{The \isafetyNo digital-safety practices we identified in our analysis of \paperNo existing works. The practices are organized into broad categories to aid readability \edit{and paired with a random sample of (up to 10) example works.}} 
\label{tab:sok-ds}
\vspace{-3mm}
\end{table*}

\noindent \edit{Researchers may deploy \textit{digital-safety practices} to mitigate the risks their research may pose to at-risk participants, the at-risk users they represent, and to the researchers themselves.}
Our analysis elicited \isafetyNo distinct digital-safety practices (SP1--SP36) summarized in \tabref{tab:sok-ds}.
We discuss these according to six high-level categories (ordered roughly chronologically relative to research project timelines).

Researchers have commenced at-risk research by forming \emph{professional partnerships}, either with practitioners (external experts) or researchers (outside of the initial research team)---for guidance, further training, or recruitment (SP1, SP2).
This also included seeking \emph{external review} outside of the bounds of a traditional \edit{research} institution (i.e., local community, NGO) to address safety concerns (SP4), or incorporating at-risk users into the research team (SP3).

If utilizing direct research approaches, researchers reported being responsive to their own \emph{positionality}, or how the personal characteristics of the researcher may influence participants' safety and the data participants are willing to provide
(SP5, SP8, SP9).
These practices included justifying their methods of \emph{participant engagement}, such as conducting pilot studies or studies with proxies for at-risk participants (SP6, SP7), as well as the need for additional training or therapeutic support for at-risk users and researchers on emotive topics (SP10, SP11).

Whether conducting direct or indirect research, works often described using \emph{privacy-preserving data collection} approaches, which involved minimizing the amount of data collected (SP12--SP15) or securing the safety of data collection sites \edit{using encryption or access control} (SP16, SP17, SP20).
At-risk participants were also commonly encouraged to use protective practices in contributing to research, such as choosing safer communication modalities or using pseudonyms (SP18, \edit{SP}19).
Works that used \emph{secure data storage and processing} aimed to significantly limit access to data pertaining to at-risk users through access control (SP21), redaction (SP22), encryption (SP23), or secure processing in transit and at rest (SP24).
These practices could also extend to preserving the privacy of the research team in contexts where the research team identified a risk of being targeted by adversaries (as described in \secref{sec:risks-researchers}).

In addition, researchers also practiced \emph{researcher accountability} to at-risk users, including adapting to the particular needs of at-risk users (SP25, SP26), ensuring transparency in what data are collected (SP27, SP29), and minimizing the risk of exploitative approaches that cause further harm (SP28).
Finally, several practices identified the need for researchers to critically analyze \emph{sharing and evaluating deliverables}, such as additional steps to ensure anonymity of those involved in research (SP30--SP33, SP35, SP36) and analysing the potential for adversaries to learn more about at-risk users (SP34).

\paragraph{Challenges.}
The numerous risks posed to at-risk participants and the users they represent (\secref{sec:risks-participants}), to researchers (\secref{sec:risks-researchers}), and the digital-safety practices used in response (\secref{sec:safety-practices}) may be a daunting array of considerations for digital-safety research.
In most cases, papers in our dataset did not provide sufficient detail, if any, about why particular risks were considered and safety practices used: \edit{27.0\%} of works \edit{($n$=53)} did not report any digital-safety practices, while \edit{9.6\%} of works \edit{($n$=19)} contained only statements clarifying approval by an ethics body (such as an IRB), and another \edit{28.5\%} of papers \edit{($n$=56)} contained one or two sentences pertaining to safety (excluding the \edit{14} works that only reported approval by an ethics body).
We determined that only \edit{32} works \edit{(16.3\%)} in our corpus contained clear justifications on why safety practices were used, who performed such practices, and actionable descriptions of the practices they employed. 
While we organised these safety practices into high-level descriptive categories (\tabref{tab:sok-ds}), these labels are not practical, actionable, or comprehensive enough to cover research from start to finish.
\section{Methods: Development of strategies}
\label{sec:methods-strategies}

\noindent Our analysis of existing works did not provide an answer to \textbf{Q3}, as pragmatic safety guidance for planning, executing, and sharing at-risk research was neither offered by nor the focus of those works. We had to go beyond the contents of those papers to gain an understanding of current practices.

\paragraph{Expert panel.}
To do so, we engaged eight S\&P scholars \edit{(all co-authors on this paper)}: established researchers with extensive experience across a broad cross-section of at-risk research.
We recruited these scholars by word-of-mouth.
The resulting \expertNo members of our expert panel (i.e., eight experienced S\&P scholars, four from our analysis team) together have over 60 years of experience in computer security and digital-safety research involving at-risk users across industry, academia, and non-profit sectors.
Our expert panel has, in aggregate, worked on many styles of research projects, ranging from short-term (less than one year) exploratory studies to long-term engagements (5~years or more) involving survivors of IPV \edit{\cite{freed_digital_2017, freed_stalkers_2018, havron2019clinical, freed_is_2019, tseng_care_2022, bellini_so-called_2021}}, political campaign workers \edit{\cite{consolvo_why_2021}}, refugees \edit{\cite{shoemaker_identity_2019}}, survivors of human trafficking \edit{\cite{chen_computer_2019}}, political activists \edit{\cite{daffalla_defensive_2021}}, journalists \edit{\cite{warford_strategies_2021}}, low-income communities \edit{\cite{woelfer_improving_2011}}, returning citizens (post-incarcerated individuals) \edit{\cite{bellini_choice-point_2020, bellini_mechanisms_2020}}, sex workers \edit{\cite{mcdonald_its_2021}}, people experiencing homelessness \edit{\cite{woelfer_homeless_2010}}, targets of occupational bullying and harassment \edit{\cite{bellini_that_2018}}, online content creators \edit{\cite{thomas_its_2022}}, and more.
We have therefore worked with a substantial cross-section---18 or 58.1\%---of the 31 at-risk groups identified in Warford et al.'s \cite{warford2022atrisk} corpus.
Our fields span S\&P, HCI, and criminology.

\paragraph{Process.}
The analysis team worked with the S\&P scholars to develop a set of strategies that would provide pragmatic guidance for the planning, execution, and sharing of digital-safety research involving at-risk users.
The nine-month process of strategy development involved three phases.

In the first phase, we conducted a series of informal discussions about safety challenges and practices in at-risk research.
In the second phase, we performed a structured oral history elicitation, known for providing a rich image of past work \cite{balaam_emotion_2019}.
The first author implemented a protocol \edit{where each expert was paired with another expert to share oral histories.
Each expert had extensive experience and training in discussing challenging topics, and were paired with experts of similar professional experience to minimize potential shame or embarrassment.
These pairings collaboratively recorded relevant experiences across their work with at-risk users in a shared, access-controlled document.}
In this way, we elicited new knowledge, not previously identified in our analysis of existing works: research experiences about digital-safety risks, which we present in this work as \textit{anecdotes}.
In total, the 12 members of the expert panel contributed 57 individual accounts of oral histories (per author M: 6.3, SD: 2), 
totalling 6,600 words (word count per item ranged from 71 to 504, M: 186, SD: 86).

We then used these oral histories in a third phase: consensus building.
\edit{The first author conducted a content analysis to identify salient themes. The expert panel regrouped to discuss these oral histories in a series of focused meetings to iterate on designing strategies for safer at-risk group research. The panel grounded the strategies in specific, actual research practice from the oral histories, making iterative refinements until we reached consensus.}
By discussing and identifying areas of disagreement, we were able to concretize six strategies for safer digital-safety research involving at-risk users.

Strategies were chosen as a viable deliverable as they necessitate a plan of action that is designed and (ideally) implemented to achieve a goal (thereby answering \textbf{Q3}).
We intend for these strategies to help researchers prioritize and customize practices appropriate to their particular research context.
The strategies, labeled \edit{S}1--\edit{S}6, appear in \tabref{tab:seven-strategies}.
We subsequently linked these strategies back to the findings from our analysis of existing works (\tabref{tab:sok-ds}), identifying how reported safety practices could be applied to support the strategies.
Note that our linking between the strategies in \tabref{tab:seven-strategies} and practices in \tabref{tab:sok-ds} \edit{are examples and not exhaustive.
The} specific practices researchers should apply for each strategy will vary depending on their research goals, context, and the people involved.

\paragraph{Limitations.}
In sharing oral histories and developing strategies, the 12 members of the expert panel provide personal accounts, which were susceptible to cognitive biases---where a researcher's expectations, opinions, prejudices, or memory may affect their ability to accurately report.
During the elicitation phase, all accounts were collected via a standardized, structured procedure and corroborated by multiple co-researchers (all co-authors on this paper) to triangulate these accounts.

Also, the six proposed strategies may not cover all instances that could affect the digital safety of participants, the group(s) they represent, or researchers during the planning, execution, and reporting of at-risk research. There will undoubtedly be instances in which a strategy does not apply or would not be appropriate.
As a result, we envision researchers using these strategies in conjunction with careful reflection on their particular context.

\section{Strategies for safer at-risk research}
\label{sec:six-strategies}
\noindent Here we present six strategies that guide researchers to think through which safety practices may apply (including but not limited to the \isafetyNo from our review) and how to enact safer research in their context.
Rather than sorting through previously applied practices---not all of which may apply to a given research context---the \strategyNo strategies raise issues that apply across many at-risk research contexts.

Our six strategies begin by highlighting a foundational frame that was omnipresent throughout our discussions, and was agreed upon by all 12 members of the expert panel to provide an orientating mindset for the six strategies:
\begin{quote}
    \emph{Research should be treated as an intervention.}
\end{quote}
This means that one should assume a priori that research will have an impact on the at-risk participants and potentially other members of the at-risk group(s) they represent.
Explicitly interventionist research traditions (e.g., clinical trials) are well-understood as sites for potential positive and negative effects, and have established procedures for handling impacts.
In digital-safety research, it is tempting to assume many types of studies---such as observational studies or online measurement studies---can be executed without affecting participants or the populations they represent.

But, as reflected in our anecdotes, even \ds research with observational or descriptive aims (e.g., surveys) can result in harm.
For example, asking participants to recount difficult experiences may trigger trauma responses in participants (\textit{re-traumatization}, \secref{sec:risks-participants}), or burnout in researchers over time (\textit{burnout and vicarious trauma}, \secref{sec:risks-researchers}).
Even online measurement studies can have impact if, for example, adversaries learn new tactics from the results, or target researchers they may who disagree with the study findings (\textit{harassment/intimidation}, \secref{sec:risks-researchers}).

After realizing that research should be treated as an intervention, the importance of employing a strategic approach should become apparent, as should the need to identify areas where strategies could be applied to mitigate risk.
We posit that planning research to minimize potential harm and maximize potential benefit is especially important when it involves at-risk users, because harm in this space has the very real potential to be outsized.

\begin{table*}[!t]
\footnotesize
\renewcommand{\arraystretch}{1.4}
\begin{tabular}{@{}p{0.3cm}p{3.7cm}p{8.4cm}p{4cm}@{}}
\toprule
\CC{40} \textbf{ID} &
  \CC{40} \textbf{Strategy title} &
  \CC{40} \textbf{Description} &
  \CC{40} \textbf{Example digital-safety practices} \\ \midrule
\edit{S}1 &
    Engage experts early &
  Consult or partner with domain experts from the beginning to inform and help facilitate safe research plans. &
  
  SP1, SP2, SP3, SP4, SP10 \\
\CC{15} \edit{S}2 & \CC{15}
  Assess and mitigate risks by threat modeling & \CC{15}
  Apply the S\&P practice of threat modeling to research protocols, and continuously update threat models to guide ongoing safety mitigations. & \CC{15}

    SP11, SP16, SP17, SP20 \\
 \edit{S}3 & 
  \edit{Select the lowest risk method that addresses the research goals} &
  Before soliciting at-risk users for high-touch methods like interviews, consider proxies (e.g., advocates), or indirect methods (e.g., online measurement). &

  SP6, SP7, SP12, SP14, SP15, SP27 \\
\CC{15} \edit{S}4 & \CC{15}
  Respect that at-risk users self-manage risk & \CC{15}
  At-risk users are often experts in managing their safety risks. Give them choice in how they engage with research safety protocols, and respect the choices they make. & \CC{15}

    SP9, SP18, SP19, SP25, SP26, SP29 \\
 \edit{S}5 &
  Be an advocate for at-risk users' needs &
  Research, by its nature, can be extractive. Build reciprocity with at-risk users, and work to help them achieve their goals. &
  SP5, SP8, SP13, SP28, SP36 \\
\CC{15} \edit{S}6 & \CC{15}
  Handle data and publications carefully & \CC{15}
  Data collection and analysis should follow security best-practice, and publications should avoid revealing identities or informing adversaries. & \CC{15}
    SP21, SP22, SP23, SP24, SP30, SP31, SP32, SP33, SP34, SP35 \\
  \bottomrule
\end{tabular}
\vspace*{2mm}
\caption{Six strategies (\edit{S1--S6}) for safer digital-safety research involving at-risk users. Each strategy represents a way of thinking about at-risk research. 
We list example safety practices for each strategy (SP1-SP36, see \tabref{tab:sok-ds}), but note that the practices researchers should apply will depend on their research goals, context, and people involved.}
\label{tab:seven-strategies}
\vspace{-3mm}
\end{table*}
\subsection{Engage experts early}
\label{sec:1-early-engagement}
\noindent Digital-safety research involving at-risk users often benefits from a wide range of expertise.
We suggest engaging experts as early as possible in research planning. They can make critical contributions to identifying appropriate safety practices and helping to ensure that potential problems are caught and corrected before they lead to harm. Early engagement also helps to avoid putting the expert(s) in the awkward situation of pointing out problems after a research protocol has been fully developed or worse, deployed.

We use ``experts'' to mean professionals or advocates who have worked with members of the at-risk group, as well as people with expertise relevant to the research more broadly. This might include  lawyers, psychotherapists, security engineers, or others well-versed in specific domains.
Moreover, users who were formerly at elevated risk of digital-safety threats (i.e., no longer under immediate threat) may be able to offer important expertise.
This strategy aims to mitigate the risk that well-intentioned---but unprepared---researchers could inadvertently conduct research that causes harm.

\paragraph{Anecdotes.}
Our work has greatly benefited from expert engagement. In an ethnographic
project with people who had experienced
incarceration \edit{\cite{bellini_mechanisms_2020}}, an early partnership with a frontline service
organization gave us access to experts who reviewed our research protocol and advised on appropriate language to
use with participants.
As researching incarceration can incur emotional responses, the organization also provided participants and the research team with therapeutic support throughout the study to help process exposure to upsetting accounts of trauma and discrimination.

In other research involving people experiencing homelessness \edit{\cite{woelfer_homeless_2010}}, our research plans were reviewed by professionals from partner support organizations.
These experts helped provide us with an overview of participants' technology use and digital-safety concerns that informed threat modeling (introduced in strategy~A2, \secref{sec:3-mitigate-risks}). 
With their help, we adapted our recruitment procedures to minimize coercion, provided safe and comfortable locations for interview sessions, and ensured ethical incentive amounts were provided.

\paragraph{Applying the strategy.}
The inclusion of experts can be useful in overseeing most, if not all, security practices (\tabref{tab:sok-ds}), yet we focus on those most relevant to experts.
Many at-risk groups have advocates and other support
professionals who might be \emph{potential research partners} (SP2) or provide \emph{external review} (SP4) of the safety of research engagements (e.g., review research protocols for safety practices).
Researchers should consider \emph{engaging with
domain experts} (SP1, SP3) from the beginning and structure the engagement to be mutually beneficial (discussed further in \edit{S}5 \secref{sec:6-advocate-need}).
Doing so helps to ensure that the research plan explicitly considers predictable effects of the research on participants (like those covered in \secref{sec:review}), and plans for making outcomes beneficial, rather than harmful.

Domain experts might be able to help the research team throughout the research process. For example, they might help think through threat models and risk mitigation, review research protocols, recruit participants or contribute to other logistics, review manuscripts for information that might identify a participant or educate an adversary (see also \edit{S}6,
\secref{sec:7-publishing-sharing}), perform or assist with direct data collection, be on call to help address unexpected situations, and more.
Domain experts can help prepare researchers for emotional reactions to the discussion of sensitive or traumatizing experiences and help plan for or even provide trauma-informed care \cite{chen_trauma-informed_2022}.
This is known as care planning, which may involve, for example,  ensuring \textit{professional therapeutic support} (SP10) is available before, during, or after the research.

\edit{All this} can involve substantial time and energy from domain experts, \edit{and so} researchers should consider how to provide proportional reimbursement---which may not necessarily be financial (see \edit{S}5, \secref{sec:6-advocate-need}).
Example approaches include paying \textit{domain experts} for their work (SP1), providing assistance to organizations via \textit{professional partnerships} in return for expert time (SP2), offering co-authorship of academic papers (SP3), or creating and disseminating research reports that would be useful for the experts and their communities.

\subsection{Assess and mitigate risks by threat modeling}
\label{sec:3-mitigate-risks}

\noindent As the name implies, digital-safety research works in an environment that can contain or attract potential adversaries.
This strategy suggests applying
threat modeling to the research process itself.
In S\&P, threat modeling is the practice of identifying relevant adversaries and enumerating their capabilities and goals (c.f.,~\cite{van2021computer}).
Identifying threat models is often the first step that engineers \edit{take to} incorporate security into system design.
We argue that threat modeling can help researchers identify and mitigate potential digital-safety risk to their participants, the group(s) they represent, or the researchers themselves.

\paragraph{Anecdotes.} Threat modeling improved the digital safety of our research involving groups concerned about surveillance by nation-state actors \edit{\cite{daffalla_defensive_2021}}.
As part of that work, we studied the privacy practices of political activists who campaigned against their government.
We began our research planning by building an understanding of threats the activists faced, in close consultation with an expert (i.e., a researcher who had personal experience in the country; see also \edit{S}1, \secref{sec:1-early-engagement}).
We determined that the activists' primary adversaries were nation-state actors who had purview over the country's entire telecommunications infrastructure, including activists' access to the global Internet.
These adversaries could arrest activists, confiscate their devices, surveil them, and even cause physical harm.
As the political climate in the country evolved, so did activists' threat models.

As a result, \edit{the interview protocol was designed to allow participants to reveal or not reveal the specific tactics they had used to evade arrests or surveillance.
The researchers in this work anticipated that some participants might need to use some of the tactics again in the future (e.g., they do not want their adversaries to discover those tactics from the research).}
In an effort to create a safe data collection environment for participants, we ensured that the researcher who conducted the interviews had a shared cultural context and spoken language with participants.
Recruitment protocols were not revealed, even in the eventual publication.
These safeguards helped alleviate participant concerns and mitigate potential harm.

\paragraph{Applying the strategy.}
Early on, we suggest that researchers perform a threat modeling exercise. This can be a brief, five-part description of (1) the adversary(ies) and their
capabilities, (2) the adversary's target and the target's defensive capabilities, (3)
the adversary's goal\edit{(s)}, (4) the impact of a successful attack (on the target or
others), and (5) the likelihood of an attack.
Even coarse estimates (e.g., likely or unlikely) can be helpful.
Most of the safety practices in Table \ref{tab:sok-ds} could be used to help assess or mitigate research harm; therefore, we discuss the most salient practices as examples to illustrate what this could encompass.

For researchers who are not very familiar with the at-risk group or context, threat modeling can be performed with the help of domain experts (see \edit{S}1, \secref{sec:1-early-engagement}) and literature reviews (see \edit{S}3, \secref{sec:4-delay-direct}).
Exploring at-risk frameworks \cite{warford2022atrisk} and systematic literature reviews \cite{sannon_privacy_2022} may also be helpful in identifying common contextual risks to consider.

Crucially, researchers should consider if and how the research might change the threats.
For example, if the threat modeling suggests
that the research might facilitate attacks or exacerbate harms, we would recommend that researchers revise their plan to include appropriate mitigations.
For instance, the threat modeling might reveal that researchers consider \textit{isolating potential threats} from at-risk groups through careful method design (e.g., separating family members to elicit data on S\&P practices \cite{mentis_upside_2019}) (SP20), implementing \textit{an anti-stalking protocol} for researchers who suspect they may be followed \cite{marczak_social_2017} (SP16), or outlining approaches to \textit{strict data access control measures} if the threat of malicious insiders becomes apparent (SP17).

Depending on the circumstances, researchers may find that the threat model does not materially change in light of the research.
Independent of what the threat modeling reveals, it offers a structured rationale to help ensure reasonable safety practices have been considered for the participants, those they represent, and the researchers.
The ability to model threats can still be continuously improved through further {digital-safety training} (SP11).

\subsection{\edit{Select the lowest risk method that addresses the research goals}}
\label{sec:4-delay-direct}

\noindent To give researchers the space to do the preparatory work to improve research safety, we suggest they
\edit{take extra time} to learn more about the at-risk group and their relevant threat models (see \edit{S}2, Section \ref{sec:3-mitigate-risks}). 
\edit{Typically the highest-risk method (aside from entirely ignoring at-risk user needs) is to directly engage at-risk users as part of the research.
However, it may also be possible to learn what is needed in lower-risk ways that avoid direct engagement (e.g., indirect measurement or proxy studies)}

\edit{Taking extra time and considering whether direct engagement may be avoided can} reduce the burden imposed on those who are already vulnerable \cite{dombrowski_social_2016}. 
This may help avoid the re-traumatization of participants or vicarious trauma for the researchers (see  Table \ref{tab:risks}). 
\edit{Any such decision should balance safety with the risk of further marginalizing an at-risk group. 
It can be harmful to de-prioritize giving marginalized groups the opportunity to speak for themselves \cite{costanza2020design}, particularly if they often have others speak on their behalf (e.g., disabled people, children \cite{mentis_upside_2019, page_perceiving_2022, spiel_adhd_2022}).
For instance, at-risk users may actively distrust the intentions of other people claiming to speak on their behalf (e.g., sex workers \cite{strohmayer_technologies_2019, bhalerao_ethics_2022}), especially if their perspectives have previously been misrepresented in published works (see misrepresentation, \secref{sec:risks-participants}).}

\paragraph{Anecdotes.} We sought to study the perspectives of abusers in technology-mediated IPV to complement work that explored survivors' perspectives. 
However, we were unsure how to safely engage abusers.
We delayed direct data collection, and instead, conducted measurement studies of  online communities that discuss technology abuse tactics \edit{\cite{tseng_tools_2020, bellini_so-called_2021}}.
Our measurement studies shed light on abuser perspectives, providing us with valuable expertise.

In our study of political activists \edit{\cite{daffalla_defensive_2021}}, we recruited ``diaspora activists,'' {proxies} for the activism movement who were from the larger at-risk group, but physically more remote from the threat.
This helped us answer our research questions, while reducing the risk of research harm.

\paragraph{Applying the strategy.}
\edit{Deciding upon the lowest-risk method to address the research goals is challenging and involves thinking through various factors, such as:
(a) Do the researchers have the experience to proceed with the method safely?
(b) What is this population's history of exclusion, and can the research be structured to not further their marginalization?
(c) Can the research risks for this method be mitigated? and 
(d) Do the benefits of using this method for the research substantially outweigh the risks?}

We suggest that before researchers determine that they must involve at-risk participants in direct engagements like interviews or focus groups, that they first explore safer, alternative methods for answering their research questions.
For example, researchers might identify \textit{proxies} (e.g., advocates who work with at-risk groups \cite{freed_digital_2017, sambasivan_privacy_2018}, people who were previously part of the at-risk group), perform pilot studies with proxies or general users, or leverage indirect data sources (e.g., public datasets, online forums, data from prior research, or academic literature) (SP6, SP7, SP27).

\edit{When marginalization is a concern, researchers can look for proxies who are closer to the population of interest, or explore alternate ways to include at-risk groups in the research that offer greater benefit to them (such as employment on the research team, SP4).}
Another approach could be to use measurement studies or sources of indirect data, such as examining online records
\cite{xue_right_2016}).

Importantly, working with proxies or indirect methods does not eliminate the need for safety mitigations or avoiding unnecessary burdens on participants.
Proxies may themselves be at-risk (as identified by \cite{masaki_exploring_2020}) and their time is valuable (see also \edit{S}5, in \secref{sec:6-advocate-need}); researchers using indirect methods should consider that some scholars have highlighted \cite{vitak_ethics_2017,pater_no_2022} that public data are not necessarily expected to be used in research.


If research with proxies or indirect methods is inappropriate, researchers could moderate the amount of data collected from at-risk groups.
This can mean \textit{discouraging participants from disclosing} sensitive information (about themselves or others \cite{venkatasubramanian_exploring_2021}), recognizing that some participants may benefit from sharing (SP12).
A commonly safety practice is to collect little-to-no \textit{identifiable information} such as participant \textit{demographics} (SP14, SP15).

\subsection{Respect that participants self-manage risk}
\label{sec:5-managing-risk}
\noindent At-risk users can be well aware of the
risks they face and active in managing these risks themselves.
To be safer \textit{and} respectful, at-risk participants should be offered the information, and authority to make decisions regarding how they will engage with safety measures planned by researchers, as part of maintaining their own safety. Researchers should plan safe options, provide information that could impact participant decisions, and guide participants in cases where they are unsure or ask for help.

\paragraph{Anecdotes.} Many of our studies embed choice in how participants engage with research across the data collection pipeline.
For example, a key concern in our research has been ensuring participants have choice over
communication modalities, since at-risk groups may be at heightened risk of surveillance. In one example, when interviewing marginalized groups (LGBTQIA+ people, women, racial/ethnic minorities) who work in computer security, to solicit sensitive anecdotes about their experiences with vulnerability discovery, we provided participants a choice of using phone calls, video chat, or voice chat on a range of platforms~\cite{fulton2023vulnerability}.
In this and other studies, we have offered participants choice around how they are represented in our data, such as allowing them to opt into audio and/or video recording, or note-taking. In nearly all cases, we offer participants the option of remaining anonymous throughout their interactions with researchers and/or in eventual publications.

This strategy extends to giving at-risk participants respect and decision-making power in their interpersonal interactions with researchers.
In our research regarding the privacy of low-income women in deeply patriarchal contexts, we learned that some women who wanted to participate feared repercussions from their husband or mother-in-law.
Thus, we offered participants the choice of speaking with us alone, in a group with other women participants, or with their husband or mother-in-law present (to mitigate the risk of harm were they to be excluded).

\paragraph{Applying the strategy.}
We suggest researchers consider how to provide options that enable participants to manage their own risks. This typically involves providing (a) choices in how participants can engage in the research, and (b) information to inform such choices.
Providing participant choice could take the form of offering multiple informed consent options or \textit{communication modalities} such as text-based or voice-based contributions \cite{sannon_privacy_2019}, as detailed above (SP19).
This may increase the complexity of studies as it relies on being \textit{responsive to emergent threats} (SP9); for example, adding remote data collection procedures may introduce threats not relevant to in-person data collection at workplaces~\cite{khovanskaya_bottom-up_2020} or public locations~\cite{marczak_social_2017}.
Complexity can also be problematic if the options overwhelm participants.

Nevertheless, we have found that even simple options and information can be beneficial---such as permitting participants to \textit{contribute false information} (e.g., a pseudonym as in \cite{mcdonald_its_2021, barwulor_disadvantaged_2021}) or for researchers to \textit{provide formal proof of identity} to potential participants (SP18, SP26).
Participants enacting choices around risk can be subtle, for example, a participant proposing \textit{data collection sessions around their schedule}, when they feel safe and able to participate (SP25).
Researchers can also be \textit{transparent about the risks of the research} (e.g., Tseng et al.'s approach to understanding escalation \cite{tseng_digital_2021}), ensuring participants have good information upon which to base their risk assessments (SP29).

This strategy does not override the need for researchers to maintain up-to-date threat models (see \edit{S}2, \secref{sec:3-mitigate-risks}): researchers should not burden participants with all threat modeling and risk assessment.
Instead, this strategy suggests that once a threat model has been identified and mitigations have been created, participants should be informed and given choice about how to engage with the mitigations---and should be trusted to make that choice.
\subsection{Be an advocate for participants}
\label{sec:6-advocate-need}
\noindent Research involving at-risk users will have an impact on them. This strategy encourages researchers to be advocates for participants, shaping beneficial impacts in the near- and long-term.
While ethical practice guides assert that the benefits of a study should outweigh the risks~\cite{protections_ohrp_belmont_2010,kenneally_menlo_2012}, this strategy goes further, encouraging researchers to think about how to structure research to have direct, near- and long-term digital-safety or other benefits for participants.

\paragraph{Anecdotes.}
We have engaged in a range of activities to advocate for and support at-risk participants in our research.
In some of our more mature work (i.e., after several years of working with the at-risk community), we have developed programs in which team members are trained to provide direct, individual assistance to at-risk individuals regarding their digital-safety \edit{\cite{havron2019clinical, freed_is_2019, tseng_care_2022, tseng_digital_2021}}.
This service enables data collection for research, but it runs as a service first and foremost.
At-risk individuals receive assistance regardless of whether they consent to participate in research, and research data is only collected if they consent.

Of course, it may not be feasible or advisable to provide such high-touch assistance.
Other ways we have advocated for or supported at-risk groups in our research include direct financial incentives (as is often suggested for work with at-risk groups~\cite{bhalerao_ethical_2022}); and collating curated lists of well-established digital-safety tools (e.g., encrypted messaging apps) or practices (e.g., using a password manager; providing pointers to advice guides that have been vetted by security experts) \edit{\cite{freed_is_2019}}.
We have also advocated for systemic change by supporting others who advocate for change, by writing and making public reports for community members, by and pushing for regulation in legislative efforts \edit{\cite{cuomo_tecc_2022}}.

\paragraph{Applying the strategy.}
To achieve immediate positive impact for participants and the group(s) they represent, researchers should adopt a mindset of being advocates through \textit{proportional incentives} (SP28).
In some cases, researchers may not know what participants want or need, especially if they are new to working with the at-risk group.
Nonetheless, we propose that by \textit{\edit{ensuring that participants engage researchers they identify with}} (e.g., \edit{they are similar in race, ethnicity,} age, gender expression) during data collection, the benefits of disclosing challenging accounts of risk (like racism to someone who ``gets it'' \cite{to_they_2020}) may be more immediately evident to the participant. (SP8)
By \textit{building rapport with participants and experts} (SP5), researchers can better understand community priorities and goals for participating in digital-safety research, and plan and implement activities that meet those needs.
This could be done by focusing \textit{data collection on supporting participants' needs} over `interesting' content from a research perspective (e.g., sensationalist content \cite{scheuerman_safe_2018, owens_you_2021}) that may not effectively advocate for their wellbeing (SP13).

\edit{Similarly, researchers could consider keeping a regular dialogue with those who work with at-risk groups to ensure they do not overpromise and underdeliver.}
While doing so is undoubtedly challenging, providing \textit{participants control of their contributions to any research deliverables} (e.g., through redaction) can help to ensure these efforts for advocacy are reflective of participant need (SP36).

This approach can be applied to shorter- and longer-term research projects; researchers should not need to approach every project involving at-risk users as though they are entering a long-term commitment.
By centering the needs of at-risk participants and the group(s) the represent, and taking reasonable steps \edit{to address them}, researchers can adopt a ``scientist-advocate'' viewpoint that can improve
research \textit{and} benefits for at-risk user(s).

We discourage researchers from promising or implying assurances that they cannot guarantee. 
For example, we suggest that researchers do not assure participants that the research will result in systemic change, or that such change will be swift. 
Effective methods to address complicated digital-safety issues are often slow to materialize, can involve unexpected road blocks, and may even depend on fundamental changes to society~\cite{sultana2018design}.
\subsection{Handle data and publications carefully}
\label{sec:7-publishing-sharing}
\noindent Our final strategy is to handle the data collected from at-risk research and the resulting
publications, talks, and other outputs with care.
While care is always recommended for human-subjects research, the sensitivity of the data (including audio or video recordings, and intermediary analysis documents) and results generated from \ds research involving at-risk users warrants special protections.

\paragraph{Anecdotes.} When we interviewed people involved with
U.S. political campaigns \edit{\cite{consolvo_why_2021}}, the audio recordings we collected contained stories that, if made public or accessed by adversaries, could potentially harm the participant's career, the campaign for which they worked, or even the political party their candidate represented.
To mitigate this risk, two research team members transcribed the audio recordings themselves to avoid sharing data outside of the immediate research team (e.g., with a professional transcription company).
Security experts reviewed our reports to ensure they did not include attacks or vulnerabilities that might inform adversaries (in this case sophisticated nation-state actors).

Across other published works, we used many other protections, including omitting details about research procedures (e.g., recruitment methods, see \edit{S}2 \secref{sec:3-mitigate-risks}) \edit{\cite{daffalla_defensive_2021}}, editing quotations \edit{\cite{freed_stalkers_2018, bellini_so-called_2021, tseng_tools_2020}}, and excluding demographic information to  prevent re-identification \edit{\cite{matthews_stories_2017, consolvo_why_2021, mcdonald_its_2021}}.

Our experiences have also demonstrated a role for caution in handling media attention after publication.
After publishing our work understanding abuser tactics in IPV, we received inquiries from reporters seeking our expertise.
Some reporters may be incentivized to seek sensationalist headlines.
For example, several have contacted us wanting to explicitly write stories on stalkerware, despite the fact that our research \edit{suggested} that it is a  less prevalent attack vector in IPV than everyday privacy violations like account compromise.
Our statements to reporters explicitly state this, and we always tell them to contact the communications staff of our partner organizations.

\paragraph{Applying the strategy.}
We suggest that
researchers map out the expected lifecycle of data collected, think through ways in
which the data may be exploitable, and define policies regarding sensitivity levels based on the threat models developed with
experts \edit{(see} Sections~\ref{sec:1-early-engagement} and~\ref{sec:3-mitigate-risks}).
This can inform mitigations, such as \textit{implementing strict data access control measures} (SP21), and \textit{deletion or redaction schedules} (SP22).
Authentication and data protection should follow the state-of-the-art in computer security, such as using \textit{secure cloud or dedicated infrastructure}, and employing \textit{encryption} (SP23, SP24).

Before publication, we suggest that \textit{papers and other research artifacts be reviewed for their potential to inform adversaries} (SP34).
Researchers could \textit{reduce the granularity of demographic
information} they report by only using aggregated summary statistics, paraphrasing participant quotes to \textit{remove or change
potentially identifying word choices}, or not \textit{associating any contributions with identifiers} (SP30, SP31, SP32, SP35).
In some cases, researchers may need to \textit{withhold the origins of data}, such as websites or physical research sites (SP33).
Researchers might also ask \textit{domain and privacy
experts} to assess the possibility of participant identification (see \edit{S}1, \secref{sec:1-early-engagement}); in doing so, it may be
helpful to provide the experts with a per-participant breakdown of all data in the publication.

A researcher's concern that publications might inform future adversaries may also cause them difficulty deciding whether and when to disclose attack details.
We suggest that researchers consider disclosing attack details when the gain from disclosure is high (e.g., to push forward future mitigations) or if adversaries can discover these tactics easily (SP34).
In these situations, reporting them is unlikely to impact future attacks.

\section{Future directions}
\label{sec:discussion-intro}
\noindent Our analysis of research risks (\secref{sec:review}) and systemization of strategies and relevant safety practices (\secref{sec:six-strategies}) can help researchers plan, execute, and report on safer digital-safety research involving at-risk users, a need highlighted by other works \cite{thomas_sok_2021, warford2022atrisk, baruh_online_2017,goerzen2019entanglements,bhalerao_ethical_2022,sannon_privacy_2022,antle_ethics_2017}.
But some of our strategies stand in contrast to, or generate friction with, research norms.
For example, delaying research efforts (\secref{sec:4-delay-direct}) may appear to conflict with the research community's drive for progress, action, and publication~\cite{asad_prefigurative_2019}.
Our suggestion to consider the impact that disclosing vulnerabilities might have on at-risk participants and the group(s) they represent (\secref{sec:7-publishing-sharing}) requires updating the
disclosure processes used by the security community \cite{cornejo_vulnerability_2016}.

Here, we discuss how these friction points are both practical career challenges for researchers and opportunities for collective improvement in the S\&P and HCI research communities.
To achieve this, we revisit the broader digital-safety research ecosystem (\edit{see} \secref{sec:related-intro}),
\edit{and discuss potential alterations to}
research \textit{deliverables}, study \textit{protocols}, research \textit{community-building}, and scientific \textit{funding}.

\paragraph{Reporting safety strategies in deliverables.}
\edit{Across the venues where digital-safety research is growing, publications rarely mention how digital safety was addressed during the research (\secref{sec:review}).}
The works that \edit{\textit{do} report safety strategies are valuable for research, practice, and training as the field evolves.}
Toward enabling future at-risk research, we suggest that the S\&P community normalize including a ``Safety strategies'' section in publications, potentially under a methods section.
It may describe the research's \edit{high-level approach} to safety, as well as what safety practices were employed to support participants, the group(s) they represent, and the researchers involved (\tabref{tab:sok-ds}).
\edit{The research community can encourage including these details in calls for papers, as well as in guidelines for peer review.}
\edit{A focus on safety strategies could draw more attention to safety in} the field, and allow researchers who are doing excellent work in this area to demonstrate their approach \edit{--- potentially leading to the discovery of unreported safety practices.}

\paragraph{\edit{Incentivizing safer research protocols.}}
The research community might also work towards establishing standards for safety practices.
Importantly, these standards must be flexible and context-sensitive, to account for the wide array of contingencies and study designs possible in at-risk research.
\edit{One way to achieve this context-specificity is to incentivize researchers to think critically and systematically about their safety procedures early, via concrete plans for safety within research proposals.}
Proposal reviewers could evaluate research plans for safety plans and make sure they are consistent with our strategies.
As a start, reviewers could consider checking whether suitable expert partners will be consulted (\edit{S}1, \secref{sec:1-early-engagement}).
Ethics review boards might similarly lean on our strategies to help in evaluating research protocols.
\edit{Such early consideration and dialogue with protocol evaluators will also smooth the way to downstream publication of safety strategies.}

\edit{Considering research safety early is particularly important in a funding climate where researchers are increasingly being asked to share anonymous data or make it accessible for secondary use.
Data sharing can play an important role in at-risk research, because it is difficult to acquire knowledge of digital-safety harms experienced by hard-to-reach populations, but it may also inadvertently enable their adversaries to worsen their attacks (\secref{sec:7-publishing-sharing}).}
\edit{To reconcile these potential harms and benefits, we encourage further research into the possibility of sharing data in at-risk research via improved computational tools, research procedures, and data-sharing paradigms.}

Increasing the complexity of digital-safety research protocols may place additional time demands on both researchers and at-risk users; \edit{for example, longer timelines to ethics board approvals} (\edit{S}5, \secref{sec:6-advocate-need}).
As such, more \edit{work} is needed to cultivate best practices throughout the research community, to ensure this time and effort is used effectively and properly rewarded \cite{warford2022atrisk, sannon_privacy_2022, ahmed_addressing_2016}.

\paragraph{Funding programs.}
\edit{One mechanism for cultivating and rewarding best-practice at-risk research is to improve research funding.}
\edit{Targeted grants could allow researchers to budget for the time and expertise to do this work,} particularly for academic researchers who primarily rely on federal or national grant funding (e.g., NSF, NIH, EPSRC, ARC).
Programs \edit{might} allow support for non-traditional roles, such as community coordinators who can manage relationships with partners \edit{or dedicated mental health and well-being support.}
In other fields, funders have programs specifically designed to train students in both practice and research, such as the U.S. NIH Medical Scientist Training Program~\cite{mstp}, which provides funding to train students as clinician-scientists qualified in medical practice and research.
Similar programs could be considered in computer science.

\paragraph{Building a digital-safety research community.}
Digital safety is becoming more critical as online hate and harassment threatens at-risk users \cite{thomas_sok_2021}, as tools for surveillance are increasingly normalized in consumer technologies \cite{tseng_tools_2020, bellini_so-called_2021, chatterjee2018spyware}, and as computing pervades critical domains like health, education, and finance.
To rise to these challenges, we need to increase the amount of research about technology and digital-safety problems involving at-risk users.
Yet safe versions of this research can take significant time and energy---to do threat modeling (\edit{S}2, \secref{sec:3-mitigate-risks}), develop partnerships with suitable
experts (\edit{S}1, \secref{sec:1-early-engagement}), prepare for emotional labor (\edit{S}3, \secref{sec:4-delay-direct}), and more.

Research communities can promote safe work with at-risk users by acknowledging this invisible labor and building support infrastructure (e.g., training resources and mentorship programs) \cite{balaam_emotion_2019, moncur_emotional_2013}.
To make safety-focused labor visible, institutions should give researchers time and incentives to do it \cite{masaki_exploring_2020}, and by training new researchers through workshop and conference development.
Ultimately, this requires a push for research communities to value quality over quantity of publications and research deliverables: \edit{a tradeoff we see as a chance to improve the work we do, in service of the at-risk communities we are inspired to support.}


\section{Acknowledgements}
\noindent \edit{We are deeply grateful to our shepherd and anonymous reviewers for their efforts to help improve this manuscript. 
We would also like to thank Elissa Redmiles, Rebecca Umbach, Georgina Powell, and the other participants of the Protecting At-Risk Users Workshop for their insightful contributions.
This work was funded in part by NSF Award \#1916096, as well as gifts from JPMorgan Chase. This material is based in part upon work supported by DARPA under grant HR00112010011. Any opinions, findings and conclusions or recommendations expressed in this material are those of the authors and do not necessarily reflect the views of the United States Government or DARPA.}

{
\small
\bibliographystyle{plain}
\bibliography{refs.bib}
}

\newpage 

\appendices 

\section{Meta-Review}

\subsection{Summary}
\noindent This paper presents a systematic analysis of 203 academic articles on digital safety research with at-risk users. By identifying 14 research risks and 36 safety practices, as well as consulting with 12 domain experts, the authors provide consolidated guidance for researchers in this field. The study highlights the need for more consistent reporting and suggests future areas of research on at-risk user research.

\subsection{Scientific Contributions}

\begin{itemize}
\item Provides a Valuable Step Forward in an Established Field
\item Other: Systematization of Knowledge and Recommendations for Researchers
\end{itemize}

\subsection{Reasons for Acceptance}
\begin{enumerate}
\item The paper is a well-written and organized resource that provides valuable insights and guidance for researchers interested in studying at-risk populations. Reviews deem it useful and worth sharing with students and researchers in the field.
\item The engagement of domain experts in the research process adds credibility and provides practical strategies for at-risk research. The reviews highlight this aspect, noting that it enhances the overall contribution of the paper.
\end{enumerate}



\end{document}